\input harvmac

\lref\kks{S. Kachru, J. Kumar and E. Silverstein, {\it 
Vacuum Energy Cancellation in a Non-supersymmetric String}, hep-th/9807076.} 

\lref\kso{S. Kachru and E. Silverstein, JHEP {\bf 11} (1998) 001.}

\lref\kst{S. Kachru and E. Silverstein, {\it On vanishing two loop
cosmological constants in nonsupersymmetric strings}, hep-th/9810129.}

\lref\dhvw{L. Dixon, J.A. Harvey, C. Vafa and E. Witten,
Nucl. Phys. {\bf B261} (1985) 678; \hfill\break 
Nucl. Phys. {\bf B274} (1986) 285.}

\lref\harv{J.A. Harvey, Phys. Rev. {\bf D59} (1999) 26002.}

\lref\toroidal{M. Bianchi, G. Pradisi and A. Sagnotti,
Nucl. Phys. {\bf B376} (1992) 365.}

\lref\newtor{M. Bianchi, Nucl. Phys. {\bf B528} (1998) 73; \hfill\break
E. Witten, JHEP {\bf 02} (1998) 006.}

\lref\bgmn{M. Bianchi, E. Gava, F. Morales and K.S. Narain, {\it D-strings in
unconventional type I vacuum configurations}, hep-th/9811013.}

\lref\kakutor{Z. Kakushadze, G. Shiu and S.-H.H. Tye, 
Phys. Rev. {\bf D58} (1998) 086001.}

\lref\new{C. Angelantonj, in preparation.}

\lref\lmml{C. Vafa, Nucl. Phys. {\bf B273} (1986) 592; \hfill\break
D.S. Freed and C. Vafa, Commun. Math. Phys. {\bf 110} (1987) 349.}

\lref\ads{I. Antoniadis, E. Dudas and A. Sagnotti, 
{\it Supersymmetry breaking, open strings and M-theory}, 
hep-th/9807011; \hfill\break
I. Antoniadis, G. D'Appollonio, E. Dudas and A. Sagnotti, {\it
Partial breaking of supersymmetry, open strings and M-theory},
hep-th/9812118.}

\lref\gp{E. Gimon and J. Polchinski, Phys.Rev. {\bf D54} (1996) 1667.}

\lref\cargese{A. Sagnotti, in Carg{\`e}se 87, Non-Perturbative Quantum Field
Theory, eds. G. Mack et al. (Pergamon Press, Oxford, 1988), p. 521.}

\lref\ps{G. Pradisi and A. Sagnotti, Phys. Lett. {\bf B216} (1989) 59.}

\lref\bs{M. Bianchi and A. Sagnotti, Phys. Lett. {\bf B247} (1990) 517,
Nucl. Phys. {\bf B361} (1991) 519.}

\lref\fps{D. Fioravanti, G. Pradisi and A. Sagnotti, Phys. Lett. {\bf
B321} (1994) 349; \hfill\break
G. Pradisi, Ya.S. Stanev and A. Sagnotti, Phys. Lett. {\bf B345}
(1995) 279; Phys. Lett. {\bf B356} (1995) 230; Phys. Lett. {\bf B381} 
(1996) 97.}

\lref\zero{A. Sagnotti, {\it Some properties of open string theories},
hep-th/9509080;\hfill\break 
A. Sagnotti, {\it Surprises in open string perturbation theory},
hep-th/9702093; \hfill\break
C. Angelantonj, Phys. Lett. {\bf B444} (1998) 309.}
\lref\zerop{R. Blumenhagen, A. Font and D. L\"ust, {\it Tachyon-free
orientifolds  of type 0B strings in various dimensions}, hep-th/9904069.}

\lref\vw{C. Vafa and E. Witten, Nucl. Phys. Proc. Suppl. {\bf 46} (1996) 225.}

\lref\bg{R. Blumenhagen and L. G\"orlich, {\it Orientifolds of
Non-Supersymmetric Asymmetric Orbifolds}, hep-th/9812158.}

\lref\st{G. Shiu and S.-H.H. Tye, {\it Bose-Fermi Degeneracy and Duality in
Non-Supersymmetric Strings}, hep-th/9808095.}

\lref\nopen{A. Dabholkar and J. Park, Nucl. Phys. {\bf B477} (1996) 701;
\hfill\break
C. Angelantonj, M. Bianchi, G. Pradisi, A. Sagnotti and Ya.S.
Stanev, Phys. Lett. {\bf B387} (1996) 743.}

\lref\lsw{W. Lerche, A.N. Schellekens and N.P. Warner,
Phys. Rept. {\bf 177} (1989) 1.}

\lref\abk{I. Antoniadis, C. Bachas and C. Kounnas, Nucl. Phys. {\bf
B288} (1987) 87; \hfill\break 
I. Antoniadis and C. Bachas, Nucl. Phys. {\bf B298} (1988)
586;\hfill\break
H. Kawai, D.C. Lewellen and S.-H.H. Tye, Nucl. Phys. {\bf B288} (1987)
1.}

\lref\nsv{K. S. Narain, M.H. Samardi and C. Vafa, Nucl. Phys. 
{\bf B288} (1987) 551.}

\lref\kk{E. Kiritsis and C. Kounnas, Nucl. Phys. {\bf B503} (1997) 117.}

\lref\ggsm{A. Sagnotti, Phys. Lett. {\bf B294} (1992) 196; \hfill\break
S. Ferrara, R. Minasian and A. Sagnotti, Nucl. Phys. {\bf B474} (1996)
323;\hfill\break
S. Ferrara. F. Riccioni and A. Sagnotti, Nucl. Phys. {\bf B519} (1998)
115; \hfill\break
F. Riccioni and A. Sagnotti, Phys. Lett. {\bf B436} (1998) 298.}
\lref\ab{I. Antoniadis and C. Bachas, {\it Branes and the Gauge Hierarchy},
hep-th/9812093.}
\lref\pw{J. Polchinski and E. Witten, Nucl. Phys. {\bf B460} (1996) 525.}
\lref\a{I. Antoniadis, Phys. Lett. {\bf B246} (1990) 377.}

\lref\low{J.D. Lykken, Phys. Rev. {\bf D54} (1996) 3693; \hfill\break
I. Antoniadis, N. Arkani-Hamed, S. Dimopoulos and G. Dvali, 
Phys. Lett. {\bf B436} (1998) 263; \hfill\break
G. Shiu and S.-H.H. Tye, Phys. Rev. {\bf D58} (1998) 106007.}
\lref\ap{I. Antoniadis and B. Pioline, {\it Low-Scale Closed Strings and their
Duals}, hep-th/9902055.}
\lref\kt{Z. Kakushadze and S.-H.H. Tye, {\it Brane World}, hep-th/9809147.}
\Title{\vbox{\rightline{\tt hep-th/9904092} \rightline{CPHT-S711.0299}}}
{\vbox{\centerline{Non-Supersymmetric Type I Strings}
\vskip .1in 
\centerline{with Zero Vacuum Energy}}}
\centerline{C. Angelantonj, I. Antoniadis and K. F\"orger}

\bigskip\centerline{\it Centre de Physique Th\'eorique}
\centerline{\it Ecole Polytechnique}
\centerline{\it F--91128 Palaiseau Cedex, FRANCE}
\vskip 1in

\centerline{{\bf Abstract}}
We study open descendants of
non-supersymmetric type IIB asymmetric (freely acting) orbifolds with zero
cosmological constant. A generic feature of these models is that
supersymmetry remains unbroken on the brane at all mass levels, while it is
broken in the bulk in a way that preserves Fermi-Bose degeneracy in both the
massless and massive (closed string) spectrum. This property remains valid in
the heterotic dual of the type II model but only for the massless excitations.
A possible application of these constructions concerns scenarios of low-energy
supersymmetry breaking  with large dimensions.

\Date{4/99} 
%

\newsec{Introduction}

One of the longest outstanding problems in theoretical physics is to explain
why the cosmological constant is extremely small and possibly vanishes
after supersymmetry breaking. It was recently argued that there is
a class of non-supersymmetric type II string compactifications whose
vacuum energy vanishes to all orders in perturbation theory
due to Fermi-Bose degeneracy \refs{\kks,\kso,\kst}. 
These models are based on asymmetric orbifold
constructions of non-abelian nature, where supersymmetry is broken in a
special way, so that the vacuum amplitude can be shown to vanish up to
genus-$2$ by explicit computation \kst. This vanishing was argued to
persist in higher orders as well, while at the non-perturbative level a
cosmological constant may be generated, although it will be exponentially
suppressed in the weak coupling limit \harv.

Since these constructions were made in the framework of type II string
theories, their obvious disadvantage is the absence of non-abelian gauge
symmetries at the perturbative level.\foot{It would be interesting to
study these models at special points of moduli space and in the presence
of RR-backgrounds, where extra massless non-abelian gauge particles appear
non-perturbatively. In section 5, 
we discuss this issue using heterotic--type II
duality.} It is then natural to ask whether they admit 
open descendants that support
non-abelian gauge groups and chiral matter, 
without destroying their main property of
vanishing vacuum energy \st. A recent analysis has indeed shown that
such a generalization is possible \bg. 

In this work we perform a systematic study of the open descendants in
non-supersymmetric type IIB asymmetric orbifolds with zero cosmological
constant, using the formalism of \refs{\cargese,\bs}. 
The models we study can
be described as Scherk-Schwarz deformations of six-dimensional (6d) vacua with
${\cal N}=1$ supersymmetry that are asymmetric 
generalizations of the class studied
in \ads. In particular, we derive the full partition 
function of the type I model based
on the freely acting orbifold of \harv. A peculiar feature
of the parent type IIB vacuum is that both $R\to\infty$ and $R\to 0$
limits lead to (6d) supersymmetric theories, with
$R$ the radius of the circle used to break supersymmetry.
This is due to an exact T-duality symmetry present
even after supersymmetry breaking. T-duality is of course broken 
in the type I theory
that has a (6d) supersymmetric decompactification limit, while for
$R\to 0$ there are linear or logarithmic divergences, in the case of one or
two dimensions, respectively. In the T-dual type I$^\prime$ picture, the
divergences reflect the existence of non vanishing local tadpoles in the
directions transverse to the D-branes \ab. 
These tadpoles remain non-vanishing
for any configuration of D-branes, unlike the familiar situation for the
${\rm SO}(16)\otimes {\rm SO}(16)$ model in nine dimensions \pw.

Although broken in the closed string sector, 
supersymmetry remains unbroken
on the D-branes. This is consistent with the results of \ads, since 
using T-duality, an exact symmetry of the parent type IIB theory,
one can always break supersymmetry along a direction transverse to the
branes.
Notice, however, that unlike the constructions of \ads, where supersymmetry
is present only for the massless excitations of the D-branes, in our case
it is preserved in all mass levels.
The same property appears to be present for the massless excitations in the
gauge (non-perturbative) sector of the type of closed string models of \harv,
at enhanced symmetry points of their moduli space, as we argue by
analysing the partition function on the heterotic side.
We also study the open descendants of another class of 
non-supersymmetric type IIB
vacua based on the free-fermionic construction \abk. These models are defined
at particular points of moduli space, and thus they can not be 
continuously deformed
to higher dimensional supersymmetric models. The open sector has a gauge group
with reduced rank and, as in the previous models, unbroken supersymmetry
at all mass levels.

The present paper is organized as follows. In section 2 we review the
non-super\-symmetric type II model of \harv \ with vanishing vacuum energy.
In section 3 we derive its one-loop partition function and discuss 
its 6d limit. In section 4 we derive its open descendants 
that have ${\cal N}=2$
unbroken supersymmetry to all mass levels, and give the full genus-1 vacuum
amplitude, that receives contributions from the torus, the Klein
bottle, the annulus and the
M\"obius strip. In section 5 we study the effects of
supersymmetry breaking on the non-abelian gauge sector of the 
heterotic dual of the type II model of \harv, by analysing the corresponding
partition function.
In section 6 we derive the open descendants of a
different type IIB 4d model, with fixed values for all the internal radii,
using the free-fermionic formulation. Finally, section 6 contains our
concluding remarks.

\newsec{The orbifold generators and their algebra}

The asymmetric orbifold we will consider in this section is generated by
the following two elements \harv:
\eqn\elem{
\eqalign{
f &= \left[ (-1^4 , 1 \,;\, 1^5 ) \,,\, (0^4 , v_{\rm L} \,;\, \delta^4
, v_{\rm R} ) \,,\, (-)^{F_{{\rm R}}} \right] \,,
\cr
g &= \left[ (1^5 \,;\, -1^4 ,1)\,,\, (\delta^4 ,w_{\rm L} \,;\,
0^4 , 
w_{\rm R} ) \,,\,
(-)^{F_{{\rm L}}} \right] \,.
\cr}
}
Here the first entry inside the square brackets denotes rotations, the
second denotes shifts on the internal compactification lattice while the
third corresponds to a genuine world-sheet symmetry. Moreover, a 
semicolon separates
holomorphic and antiholomorphic coordinates. 
In order to implement an asymmetric $Z_2$ rotation, the internal  
four-dimensional torus must split into 
a product of four circles with self-dual radius 
$R = \sqrt{\alpha'}$, such that the lattice factorizes 
into a holomorphic and an antiholomorphic part.
$\delta$ is then a shift by $R/2$,
as required by level matching and multi-loop modular invariance 
conditions \lmml. 
No further constraints are imposed
on the radius of the fifth coordinate, while $v_{{\rm L},{\rm R}}=w_{{\rm
R},{\rm L}}$.  
The shifts $v_{\rm L,R}$ 
in the fifth
coordinate, $A_2$ shifts in the notation of \vw,
act as
$$
\eqalign{
X(z) &\to X(z) + {\textstyle{1\over 2}} \left({\alpha ' \over R} + R
\right)\,,
\cr
\overline{X} (\bar z) &\to \overline{X} (\bar z ) - {\textstyle{1\over
2}} \left( {\alpha ' \over R} - R \right) \,,
\cr}
$$
and give a contribution
$$
{\textstyle {1\over 8}} \left( {\alpha ' \over R} + R \right)^2 - 
{\textstyle {1\over 8}} \left( {\alpha ' \over R} - R \right)^2 =
{\textstyle {1\over 2}}\alpha ' \,,
$$
to the level matching condition, thereby balancing the contribution of the
shifts that act on $T^4$. 

Due to the presence of $(-)^{F_{{\rm L,R}}}$, the  $g$ ($f$) generator
projects out all the gravitini coming from the (anti-)holomorphic
sector, and therefore the combined action of $f$ and $g$ breaks
supersymmetry completely. The shifts give masses to the corresponding
twisted sectors, ensuring that no massless gravitons originate from them.
Although the model is non-supersymmetric,
it has been argued in \kks\ that one-loop and higher order 
perturbative corrections to the cosmological constant vanish.
There are, however, non-perturbative contributions originating from
wrapped D-branes that can be studied perturbatively on the dual 
heterotic theory \harv.

The generators in \elem\ satisfy the algebra
$$
\eqalign{
f \circ f &= \left[ (1^5 \,;\, 1^5 )\,,\, (0^5 \,;\, (2 \delta)^4 , 0) \,,\,
1\right] \,,
\cr
g \circ g &= \left[ (1^5 \,;\, 1^5 ) \,,\, ((2 \delta)^4 ,0 \,;\, 0^5 ) \,,\,
1\right] \,,
\cr
f \circ g &= \left[ (-1^4 ,1 \,;\, -1^4 , 1) \,,\, (-\delta^4 , 0 \,;\,
\delta^4 , 0) \,,\, (-)^{F_{{\rm L}} + F_{{\rm R}}} \right]\,,
\cr
g \circ f &= \left[ (-1^4 ,1 \,;\, -1^4 , 1) \,,\, (\delta^4 , 0 \,;\,
-\delta^4 , 0) \,,\, (-)^{F_{{\rm L}} + F_{{\rm R}}} \right] \,,
\cr}
$$
thus revealing the non-abelian nature of the orbifold group $S$. 
When restricted to the point group $\overline{P}$, 
defined as the quotient of
the space group $S$ by the generators of pure translations \dhvw,
the orbifold group reduces to a simple abelian group, 
$Z_2 \otimes Z_2$, where the $Z_2$ factors
are generated by $f$ and $g$. Then one can first mod out the theory
by $\Lambda = \{ f^2 , g^2 \}$, thereby defining a new compactification
lattice, and then quotient by the point group $\overline{P} = Z_2
\otimes Z_2$. In the next section we will follow this procedure. 
In fact, the model \elem \ is a freely acting asymmetric orbifold of
$T^4/Z_2$ by $f$, and can also be interpreted as an
asymmetric Scherk-Schwarz 
deformation by doubling the radius of the fifth coordinate 
\refs{\kk,\ads}. 
Due to the asymmetric nature of the deformation that acts
simultaneously on momentum and winding modes, the lattice contribution
to the resulting model is invariant under
T-duality. As a result, the standard Scherk-Schwarz breaking
model is equivalent to the M-theory breaking one.
An alternative approach, leading to the same result, would be to mod out
directly by the space group $S$ \bg.


\newsec{The torus partition function}

The starting point in the construction of the orbifold described in
the previous section is the toroidal compactification of the type II
superstring on a five-dimensional lattice $\Gamma_{(5,5)} = [
\Gamma_{{\rm SU} (2)} ]^4 \times [\Gamma_{(1,1)} (R)]$, where 
$\Gamma_{{\rm SU} (2)}$ denotes the ${\rm SU} (2)$ lattice and
$\Gamma_{(1,1)} (R)$ is the contribution of a single circle of radius
$R$. In order to construct the open-descendants of this model we
start from the type IIB superstring, that is invariant under the
action of the world-sheet parity $\Omega$. Using the ${\rm SO} (8)$
characters $O_8$, $V_8$, $S_8$ and $C_8$, associated to the
conjugacy classes of 
identity, vector, spinor and conjugate spinor representations, respectively, 
to represent the contribution
of the world-sheet fermions, the partition function reads:
\eqn\tpfsu{
{\cal T}_0 = |V_8 - S_8 |^2 \, \left[ |\chi_{{\bf 1}}|^2 + |\chi_{{\bf
2}}|^2 \right]^4 \, {\cal Z}_{m,n} (q, \bar q) \,.
}
Here $\chi_{{\bf 1}}$ and $\chi_{{\bf 2}}$ indicate the two characters
of ${\rm SU} (2)$ at level one, corresponding to the 
conjugacy classes of the singlet and the
doublet representation, and 
$$
{\cal Z}_{m,n} (q ,\bar q) = \sum_{m,n\in Z} q^{{\alpha '\over 4} (m/R +
nR/\alpha ')^2} \, \bar q ^{{\alpha '\over 4} (m/R - nR / \alpha
' )^2} \,,
$$
with $q={\rm e}^{2\pi{\rm i}\tau}$.

Following the strategy outlined in the previous section,
we mod out the toroidal amplitude \tpfsu\ by the generators of
$\Lambda$ that act as pure asymmetric lattice shifts. This results in
a new toroidal compactification where 
the ${\rm SU}(2)^4$ lattice is turned into an ${\rm SO} (8)$ lattice:
$$
{\cal T}_{\Lambda} = |V_8 - S_8 |^2 \, \left[ |O_8|^2 + |V_8|^2 
+ |S_8|^2 + |C_8|^2 \right] \, {\cal Z}_{m,n} (q,\bar q) \,.
$$
The presence of a non-trivial
background for the ${\rm NS}\otimes {\rm NS}$ antisymmetric
tensor in the ${\rm SO} (8)$
lattice reduces the rank of the Chan-Paton (CP) 
gauge group by a factor equal to the rank of the
antisymmetric tensor \refs{\toroidal,\newtor,\kakutor,\new}.
As we will see in the next section, this turns
out to be crucial in order to get a consistent model. 

Acting now with the point group $\overline{P}$, one gets the
following formal expression for the partition function 
\eqn\fpfo{
\eqalign{
{\cal T} = & {\textstyle{1\over 4}} \big[ {\cal T}_{(0,0)} + 
{\cal T}_{(0,f)} + {\cal T}_{(0,g)} + {\cal T}_{(0,fg)} + 
\cr
&\ + {\cal T}_{(f,0)} + {\cal T}_{(f,f)} + {\cal T}_{(g,0)} + {\cal T}_{(g,g)} 
+ {\cal T}_{(fg,0)} + {\cal T}_{(fg,fg)} \big] \,.
\cr}
}
Note the absence of the disconnected modular orbit generated by ${\cal
T}_{(f,g)}$. This fact has two different
explanations in the space group or in the point group
approaches of the orbifold. In
the former case, it is due to the fact that the path integral receives
contributions only from commuting pairs of spin structures \dhvw, whereas in
the latter case, ${\cal T}_{(f,g)}$ vanishes due to
the simultaneous action of shifts and rotations \kk.

The amplitudes ${\cal T}_{(a,b)}$ with $a,b\in\{0,f,g,fg\}$ are given by
\eqn\utpf{
\eqalign{
{\cal T}_{(0,0)} &= |V_4 O_4 + O_4 V_4 - S_4 S_4 - C_4 C_4 |^2 \left[
|O_8 |^2 + |V_8 |^2 + |S_8 |^2 +|C_8|^2 \right] {\cal Z}_{m,n} \,,
\cr
{\cal T}_{(0,f)} &= (V_4 O_4 - O_4 V_4 +S_4 S_4 - C_4 C_4 )
(\overline{V}_4 \overline{O}_4 + \overline{O}_4 \overline{V}_4 +
\overline{S}_4 \overline{S}_4 + \overline{C}_4 \overline{C}_4 ) \times
\cr
&\quad \times {\vartheta^{2}_{3} \vartheta^{2}_{4} \over \eta^4}
\, (\overline{O}_4 \overline{O}_4 - \overline{V}_4 \overline{V}_4) 
\, (-)^{m+n} {\cal Z}_{m,n} \,,
\cr
{\cal T}_{(0,g)} &= (V_4 O_4 + O_4 V_4 + S_4 S_4 + C_4 C_4 ) (
\overline{V}_4 \overline{O}_4 - \overline{O}_4 \overline{V}_4 +
\overline{S}_4 \overline{S}_4 - \overline{C}_4 \overline{C}_4 ) \times
\cr
&\quad \times (O_4 O_4 - V_4 V_4) \, {\overline{\vartheta}^{2}_{3}
\overline{\vartheta}^{2}_{4} \over \overline{\eta}^4} \, (-)^{m+n}
{\cal Z}_{m,n} \,,
\cr
{\cal T}_{(0,fg)} &= |V_4 O_4 - O_4 V_4 - S_4 S_4 + C_4 C_4 |^2 \,
\left| {\vartheta^{2}_{3} \vartheta^{2}_{4} \over \eta^4} \right|^2 \,
{\cal Z}_{m,n}\,,
\cr}
}
in the untwisted sector, by
\eqn\ftpf{
\eqalign{
{\cal T}_{(f,0)} &= {\textstyle{1\over 2}}  
\, (O_4 C_4 + V_4 S_4 - S_4 O_4 - C_4 V_4)
(\overline{O}_4 \overline{O}_4 + \overline{V}_4 \overline{V}_4 -
\overline{S}_4 \overline{C}_4 - \overline{C}_4 \overline{S}_4 ) \times
\cr
&\quad \times {\vartheta^{2}_{2} \vartheta^{2}_{3} \over  \eta^4}\,
[(\overline{O}_4 + \overline{V}_4)(\overline{S}_4 + \overline{C}_4 )
+(\overline{S}_4 + \overline{C}_4 )(\overline{O}_4 + \overline{V}_4)]
\, {\cal Z}_{m+1/2,n+1/2} \,,
\cr
{\cal T}_{(f,f)} &= {\textstyle{1\over 2}}
\, (O_4 C_4 - V_4 S_4 - S_4 O_4 + C_4 V_4 ) 
( \overline{O}_4 \overline{O}_4 + \overline{V}_4 \overline{V}_4 +
\overline{S}_4 \overline{C}_4 + \overline{C}_4 \overline{S}_4 )\times
\cr
&\quad \times {\vartheta^{2}_{2} \vartheta^{2}_{4} \over  \eta^4} \,
[(\overline{O}_4 - \overline{V}_4 )(\overline{S}_4 +
\overline{C}_4)
+(\overline{S}_4 + \overline{C}_4 )(\overline{O}_4 -
\overline{V}_4)]\, (-)^{m+n} {\cal Z}_{m+1/2,n+1/2} \,,
\cr}
}
in the $f$-twisted sector, by
\eqn\gtpf{
\eqalign{
{\cal T}_{(g,0)} &= {\textstyle{1\over 2}}
\, (O_4 O_4 + V_4 V_4 - S_4 C_4 - C_4 S_4 )
(\overline{O}_4 \overline{C}_4 + \overline{V}_4 \overline{S}_4 -
\overline{S}_4 \overline{O}_4 - \overline{C}_4 \overline{V}_4 )\times
\cr
& \quad \times [(O_4 + V_4 )(S_4 + C_4)+(S_4 + C_4)(O_4 + V_4)] \,
{\overline{\vartheta}^{2}_{2} \overline{\vartheta}^{2}_{3} \over 
\overline{\eta}^4}\, {\cal Z}_{m+1/2,n+1/2} \,,
\cr
{\cal T}_{(g,g)} &= {\textstyle{1\over 2}} 
\, (O_4 O_4 + V_4 V_4 + S_4 C_4 + C_4 S_4 )
(\overline{O}_4 \overline{C}_4 - \overline{V}_4 \overline{S}_4 -
\overline{S}_4 \overline{O}_4 + \overline{C}_4 \overline{V}_4 )\times
\cr
& \quad \times [(O_4 - V_4 )(S_4 + C_4) + (S_4 + C_4)(O_4 - V_4)] \,
{\overline{\vartheta}^{2}_{2} \overline{\vartheta}^{2}_{4} \over 
\overline{\eta}^4}\, (-)^{m+n} {\cal Z}_{m+1/2,n+1/2} \,,
\cr}
}
in the $g$-twisted sector, and by 
\eqn\fgtpf{
\eqalign{
{\cal T}_{(fg,0)} &= \, |O_4 S_4 + V_4 C_4 - C_4 O_4 - S_4 V_4 |^2
\, \left| {\vartheta^{2}_{2} \vartheta^{2}_{3} \over 
\eta^4}\right|^2 \, {\cal Z}_{m,n} \,,
\cr
{\cal T}_{(fg,fg)} &= \, |O_4 S_4 - V_4 C_4 - C_4 O_4 + S_4 V_4 |^2
\, \left| {\vartheta^{2}_{2} \vartheta^{2}_{4} \over \eta^4
}\right|^2 \, {\cal Z}_{m,n} \,,
\cr}
}
in the $fg$-twisted sector. Here $\vartheta_j$ are 
Jacobi theta functions, while $\eta$ is the Dedekind eta-function. 
In order to implement the orbifold
projection, we have broken the ${\rm SO} (8)$ characters into products
of ${\rm SO} (4)$ characters, and we have used the
$S: \tau\to-1/\tau$ and $T:\tau\to \tau+1$ modular transformation matrices
$$
S_{{\rm SO}(2n)} = {1 \over 2} \left( 
\matrix{
1 & 1 & 1 & 1
\cr
1 & 1 & -1 & -1 
\cr
1 & -1 & {\rm i}^{-n} & -{\rm i}^{-n}
\cr
1 & -1 & -{\rm i}^{-n} & {\rm i}^{-n}
\cr}
\right) \,,
\quad
T_{{\rm SO} (2n)} = {\rm e}^{- {\rm i} \pi n/ 12} \, {\rm diag}\, \left( 1,
-1 , {\rm e}^{{\rm i}\pi n/4} ,{\rm e}^{{\rm i}\pi n/4}\right)
\,.
$$
From expressions \utpf-\fgtpf, one can check that the partition function
vanishes identically in each sector. 

Expanding the internal characters
and the theta functions in powers of $q$, and keeping only  leading
terms, we get the following massless contributions
\eqn\tamm{
\eqalign{
{\cal T}_{{\rm untw}} &\sim |V_4 O_4 |^2 + |S_4 S_4 |^2 - (O_4 V_4 )(
\overline{C}_4 \overline{C}_4 ) - (C_4 C_4 )( \overline{O}_4
\overline{V}_4 ) \,,
\cr
{\cal T}_{fg{\rm -tw}} &\sim 8\, |O_4 S_4 - C_4 O_4 |^2 \,,
\cr}
}
that translate into the following   
five-dimensional field content: $\{ g_{\mu\nu}, 7 A_\mu , 6 \phi ; 8 \psi \}$
from the untwisted sector, and $8\, \{ A_\mu ,
5\phi ; 2 \psi\}$ from the $fg$-twisted sector. Here, 
$g_{\mu\nu}$, $A_\mu$, $\phi$ and $\psi$ denote 
the five dimensional
graviton, vector, real scalar and Dirac fermion, respectively.
The factor 8 in 
${\cal T}_{fg{\rm -tw}}$ counts the number of fixed points left
invariant by the shifts. 

Due to the asymmetric shift on the fifth coordinate, the torus
amplitude is invariant under T-duality. As a result, the amplitude
in the decompactification limit $R\to \infty$ coincides with the one 
in the $R\to 0$ limit, aside from volume factors.
The former case corresponds to the decompactification limit of type IIB
on the supersymmetric $T^4 /Z_2$ orbifold, with 21 tensor
multiplets coupled to ${\cal N}=(2,0)$ supergravity in six dimensions.
The latter case has a natural interpretation in terms of type IIA
compactified on the same orbifold, with 6d ${\cal N}=(1,1)$ supergravity
coupled to 20 vector multiplets.
In order to have a consistent assignment of
quantum numbers \ads, in the $R\to\infty$ ($R\to 0$) limit 
it is required to double (halve) the radius.
Consequently, the terms ${\cal Z}_{m,n}$ contribute with 
an additional factor of 2 that partially
compensates the $1/4$ factor in the partition function \fpfo, yielding the
expected multiplicity of states in the $T^4 /Z_2$ orbifolds.


\newsec{Open descendants}

Following \refs{\cargese,\bs} \ the construction 
of open descendants resembles a
$Z_2$ orbifold where the $Z_2$ symmetry is the world-sheet parity
$\Omega$. The ``untwisted sector'' of the parameter space orbifold
consists of closed unoriented strings, whose contribution to the
total partition function is
$$
{\textstyle{1\over 2}} ({\cal T} + {\cal K})\,,
$$
where the Klein bottle amplitude
\eqn\kbd{
\eqalign{
{\cal K} =& {\textstyle{1\over2}} \biggl[ (V_4 O_4 + O_4 V_4 - S_4 S_4 - C_4
C_4 ) \, (O_8 + V_8 + S_8 + C_8) \, {\cal Z}_m \,+
\cr
&\quad + (V_4 O_4 - O_4 V_4 - S_4 S_4 + C_4 C_4 ) \,
{\vartheta^{2}_{3} \vartheta^{2}_{4} \over \eta^4}\, (-)^m {\cal Z}_m
\, +
\cr
&\quad + 2 (O_4 S_4 + V_4 C_4 - C_4 O_4 - S_4 V_4 ) \,
{\vartheta^{2}_{2} \vartheta^{2}_{3} \over \eta^4}\, {\cal Z}_m \,
\biggr]\,,
\cr}
}
completes the $\Omega$ projection. In the following we will use the
same conventions as in \ads\ for the lattice sums
$$
{\cal Z}_{m+a} = \sum_{m\in Z} q^{{1\over 2} [(m+a)/R]^2} \,, \qquad 
\widetilde{\cal Z}_{n+b} = \sum_{n\in Z} q^{{1\over 2} [(n+b)R/2]^2} \,,
$$
that correspond to the choice $\alpha ' = 2$. 
At the massless level one finds
$$
{\cal K} \sim (V_4 O_4 - S_4 S_4) + 4\, (O_4 S_4 - C_4 O_4) \,,
$$
that symmetrizes correctly the torus amplitude \tamm. The factor 4
has to be interpreted as $n_+ - n_-=6-2$, where $n_\pm$ is the
number of fixed points with $\Omega = \pm 1$ \refs{\kakutor,\new}. 
The resulting
spectrum of massless excitations then results in the following
five-dimensional fields: $\{ g_{\mu\nu}\,,\, 2 A_\mu \,,\, 5 \phi
\,;\, 4 \psi \}$ from the untwisted sector and $\{ 2 A_\mu \,,\,
 26 \phi \,;\, 8 \psi \}$ from the $fg$-twisted sector. Since ${\cal K}$
is supersymmetric, the contribution
of the Klein bottle amplitude to the 1-loop cosmological constant
vanishes identically. Although the asymmetric nature of the orbifold
is reflected in the presence of the signs in the lattice sums in \kbd,
the open descendants only feel the left-right symmetric (supersymmetric) 
generator $fg$ \kt.

In the transverse channel, the Klein bottle amplitude is
$$
\eqalign{
\widetilde{\cal K} =& {2^4 \over 2} \, R\, \biggl[ (V_4 O_4 + O_4 V_4 - S_4
S_4 - C_4 C_4 ) \, O_8 \, \widetilde{\cal Z}_{2n} \, +
\cr
&\quad + (V_4 O_4 - O_4 V_4 - S_4 S_4 + C_4 C_4 )\, {\vartheta^{2}_{3}
\vartheta^{2}_{4} \over \eta^4}\, \widetilde{\cal Z}_{2n} \, +
\cr
&\quad + {\textstyle{1\over 2}}\, (O_4 S_4 + V_4 C_4 - C_4 O_4 - S_4 V_4 ) \,
{\vartheta^{2}_{2} \vartheta^{2}_{3} \over \eta^4}\, \widetilde{\cal
Z}_{2n+1} \, \biggr] \,,
\cr}
$$
with a massless tadpole proportional to
\eqn\kbtm{
\widetilde{\cal K} = 2^4 \, R\, (V_4 O_4 - S_4 S_4 ) + {\rm massive} \,.
}
There are no volume factors relative to $T^4$, because we are
working at the ${\rm SO} (8)$ rational point. Something
particularly interesting 
happens due to the presence of the ${\rm SO} (8)$ lattice, which
ensures the consistency of the theory. 
To appreciate the meaning of eq. \kbtm, let us recall that,
in the presence of a quantized background for
the ${\rm NS} \otimes {\rm NS}$ antisymmetric tensor $B_{IJ}$, the
transverse Klein bottle amplitude associated to the $T^4 /Z_2$ orbifold
reads \new
\eqn\tkbwat{
\widetilde{\cal K} \sim \left[ \sqrt{v} + {2^{-r/2} \over \sqrt{v}}
\right]^2 \, (V_4 O_4 - S_4 S_4 ) + \left[ \sqrt{v} - {2^{-r/2} \over
\sqrt{v}} \right]^2 \, (O_4 V_4 - C_4 C_4 )\,.
}
Here $v$ is the volume of the internal lattice and $r$ is the rank of 
$B_{IJ}$.
For the ${\rm SO} (8)$ lattice $r=2$ and $v ={1\over 2}$ \lsw , 
and the transverse Klein bottle amplitude is
\eqn\tkbso{
\widetilde{\cal K} \sim (V_4 O_4 - S_4 S_4 ) \,.
}
Although both \tkbwat\ and \tkbso\ are consistent with the
associated torus amplitude
\eqn\torb{
{\cal T} \sim |V_4 O_4 - S_4 S_4 |^2 + |O_4 V_4 - C_4 C_4 |^2 +\ldots
\,,
}
only \tkbso\ is compatible with \tamm . A similar phenomenon will take
place in the open sector, along with an identification of
Neumann and Dirichlet charges. 

The ``twisted sector'' of the parameter space orbifold
$$
{\textstyle{1\over 2}} ({\cal A} + {\cal M} )\,,
$$
corresponds to the open unoriented sector of the spectrum and carries
multiplicities associated to CP charges (D-branes) that live at the
ends of the open strings. In five dimensions, there are two different
open sectors that differ by a sign in the massive $fg$-twisted
states. This ambiguity is related to the two possible identifications
of Neumann and Dirichlet charges under the action of the $f$ and $g$
generators. The annulus and M\"obius strip amplitudes are then given by
\eqn\andp{
\eqalign{
{\cal A}_{1,2} =& {\textstyle{1\over 2}}
\biggl[ (M+\overline{M})^2 \, (V_4 O_4 + O_4 V_4 - S_4 S_4 -
C_4 C_4 )\, O_8 +
\cr
&\quad - (M-\overline{M})^2 \, 
(V_4 O_4 - O_4 V_4 - S_4 S_4 +C_4 C_4 ) \, 
{\vartheta^{2}_{3} \vartheta^{2}_{4} \over \eta^4} \biggr]
\, {\cal Z}_{2m}
\cr
& +{\textstyle{2\over 8}}\, 
\biggl[ (M+\overline{M})^2 \, (O_4 S_4 + V_4 C_4 - C_4 O_4 - S_4
V_4 )\, {\vartheta^{2}_{2} \vartheta^{2}_{3} \over \eta^4}\, +
\cr
&\quad \mp
(M-\overline{M})^2 (O_4 S_4 - V_4 C_4 - C_4 O_4 + S_4 V_4 )\,
{\vartheta^{2}_{2} \vartheta^{2}_{4} \over \eta^4}\biggr] \, {\cal
Z}_{2m+1} \,,
\cr}
}
and 
\eqn\msdp{
\eqalign{
{\cal M}_{1,2} =& - {\textstyle{1\over 2}} (M+\overline{M}) \, \biggl[
(\widehat{O}_4 \widehat{V}_4 - \widehat{V}_4 \widehat{O}_4 +
\widehat{S}_4 \widehat{S}_4 - \widehat{C}_4 \widehat{C}_4 )
\widehat{O}_8 
\cr
&\qquad + (\widehat{V}_4 \widehat{O}_4 + \widehat{O}_4 \widehat{V}_4 -
\widehat{S}_4 \widehat{S}_4 - \widehat{C}_4 \widehat{C}_4 ) \, 
{\widehat{\vartheta}^{2}_{3} \widehat{\vartheta}^{2}_{4} \over
\widehat{\eta}^4} \, \biggr] \widehat{\cal Z} _{2m} \,+
\cr
& +{\textstyle{2\over 8}}\, (M-\overline{M})\, \biggl[
(\widehat{O}_4 \widehat{S}_4 - \widehat{V}_4 \widehat{C}_4 -
\widehat{C}_4 \widehat{O}_4 + \widehat{S}_4 \widehat{V}_4 )
\,{\widehat{\vartheta}^{2}_{2} \widehat{\vartheta}^{2}_{3} \over
\widehat{\eta}^4} \, +
\cr
&\qquad \mp\, (\widehat{O}_4 \widehat{S}_4 + \widehat{V}_4
\widehat{C}_4 -\widehat{C}_4 \widehat{O}_4 - \widehat{S}_4
\widehat{V}_4 ) \, {\widehat{\vartheta}^{2}_{2}
\widehat{\vartheta}^{2}_{4} \over \widehat{\eta}^4}\,\biggl]\, (-)^m
\widehat{{\cal Z}}_{2m+1}\,,
\cr}
}
where in the M\"obius strip amplitude we have defined suitable hatted
characters \bs. The parametrization of the annulus and M\"obius strip
amplitudes in terms of CP charges $M$ and $\overline M$ follows from
the requirement of a consistent particle interpretation for the
spectrum and from the tadpole conditions, that lead also to
a supersymmetric open unoriented sector.
In the transverse channel, these amplitudes are
$$
\eqalign{
\widetilde{\cal A}_{1,2} =& {2^{-4}\over 2}\, R \, (M+\overline{M})^2 \biggl[ 
(V_4 O_4 + O_4 V_4 - S_4 S_4 - C_4 C_4 ) (O_8 + V_8 + S_8 + C_8 )
\widetilde{\cal Z}_n +
\cr
& \quad + (V_4 O_4 - O_4 V_4 - S_4 S_4 + C_4 C_4 ) \,
{\vartheta^{2}_{3} \vartheta^{2}_{4} \over \eta^4} \, (-)^n
\widetilde{\cal Z}_n \biggr] +
\cr
& - {2^{-3} \over 8}\, R \, (M-\overline{M})^2 \, \biggl[
4 \, (O_4 S_4 + V_4 C_4 - C_4 O_4 - S_4 V_4 )\, {\vartheta^{2}_{2}
\vartheta^{2}_{3} \over \eta^4}\, \widetilde{\cal Z}_n +
\cr
&\quad \pm 2 (O_4 S_4 - V_4 C_4 - C_4 O_4 + S_4 V_4 )\,
{\vartheta^{2}_{2} \vartheta^{2}_{4} \over \eta^4} \, (-)^n
\widetilde{\cal Z}_n \biggr] \,,
\cr}
$$
and
$$
\eqalign{
\widetilde{\cal M}_{1,2} =& - {\textstyle{2\over 2}}\, R \, 
(M+\overline{M}) \, \biggl[ (\widehat{V}_4
\widehat{O}_4 - \widehat{O}_4 \widehat{V}_4 - \widehat{S}_4
\widehat{S}_4 + \widehat{C}_4 \widehat{C}_4 ) \widehat{O}_8 +
\cr
&\qquad +( \widehat{V}_4 \widehat{O}_4 + \widehat{O}_4 \widehat{V}_4 -
\widehat{S}_4 \widehat{S}_4 - \widehat{C}_4 \widehat{C}_4 ) \,
{\widehat{\vartheta}^{2}_{3} \widehat{\vartheta}^{2}_{4} \over
\widehat{\eta}^4} \biggr] 
\widehat{\widetilde{\cal Z}} _{2n} +
\cr
&+{\textstyle{4\over 8}}\, R \,(M-\overline{M})\bigg[
(\widehat{O}_4 \widehat{S}_4 - \widehat{V}_4 \widehat{C}_4 -
\widehat{C}_4 \widehat{O}_4 + \widehat{S}_4 \widehat{V}_4 )
\,{\widehat{\vartheta}^{2}_{2} \widehat{\vartheta}^{2}_{3} \over
\widehat{\eta}^4} \, +
\cr
&\qquad \mp\, (\widehat{O}_4 \widehat{S}_4 + \widehat{V}_4
\widehat{C}_4 -\widehat{C}_4 \widehat{O}_4 - \widehat{S}_4
\widehat{V}_4 ) \, {\widehat{\vartheta}^{2}_{2}
\widehat{\vartheta}^{2}_{4} \over \widehat{\eta}^4}
\bigg](-1)^n \widehat{{\widetilde{\cal Z}}}_{2n+1}
\cr}
$$
that imply the following contributions to the massless tadpoles:
$$
\widetilde{\cal A}_{1,2} =\, 2^{-4} \, R\, (M+ \overline{M})^2 \, (V_4 O_4 -
S_4 S_4 ) -2^{-2} \, R\, (M-\overline{M})^2 (O_4 S_4 - C_4 O_4 ) +
{\rm massive}\,,
$$
and
$$
\widetilde{\cal M}_{1,2} =\, - 2 \, R\, (M+\overline{M})\, (\widehat{V}_4
\widehat{O}_4 - \widehat{S}_4 \widehat{S}_4 ) + {\rm massive}.
$$
Together with the contribution \kbtm\ from the transverse Klein bottle
amplitude, tadpole conditions result in the following constraints on
the CP charges:
\eqn\tadpoles{
M+\overline{M} = 16\,, \qquad M-\overline{M} =0\,.
}

Keeping only the leading terms in the expansion of the direct 
amplitudes \andp\ and \msdp\ yields
$$
\eqalign{
{\cal A}_1 \sim & \biggl[ 2 M\overline{M} (V_4 O_4 - S_4 S_4 ) + (M^2 +
\overline{M}^2 ) (O_4 V_4 - C_4 C_4) \biggr] \, {\cal Z}_{2m} \, +
\cr
&+ 2 \, \biggl[ 2 M \overline{M} (O_4 S_4 - C_4 O_4) + (M^2 +
\overline{M}^2 ) (V_4 C_4 - S_4 V_4) \biggr]\, {\cal Z}_{2m+1} \,,
\cr}
$$
$$
{\cal M}_1 \sim  -(M+\overline{M}) (\widehat{O}_4 \widehat{V}_4 - 
\widehat{C}_4 \widehat{C}_4 ) \widehat{\cal Z}_{2m} 
+ 2\, (M-\overline{M}) (\widehat{V}_4 \widehat{C}_4 - \widehat{S}_4
\widehat{V}_4 ) (-)^m \widehat{\cal Z}_{2m+1}\,,
$$
and
$$
\eqalign{
{\cal A}_2 \sim & \biggl[ 2 M\overline{M} (V_4 O_4 - S_4 S_4 ) + (M^2 +
\overline{M}^2 ) (O_4 V_4 - C_4 C_4) \biggr] \, {\cal Z}_{2m} \, +
\cr
&+ 2 \, \biggl[ (M^2 +\overline{M}^2) (O_4 S_4 - C_4 O_4) + 2 M
\overline{M} (V_4 C_4 - S_4 V_4) \biggr]\, {\cal Z}_{2m+1} \,,
\cr}
$$
$$
{\cal M}_2 \sim  -(M+\overline{M}) (\widehat{O}_4 \widehat{V}_4 - 
\widehat{C}_4 \widehat{C}_4 ) \widehat{\cal Z}_{2m} 
+ 2\, (M-\overline{M}) (\widehat{O}_4 \widehat{S}_4 - \widehat{C}_4
\widehat{O}_4 ) (-)^m \widehat{\cal Z}_{2m+1}\,.
$$
The spectrum associated
with the open unoriented sector can thus be arranged in ${\cal N}=2$
five-dimensional supersymmetric representations, and comprises a 
vector multiplet in the adjoint representation of
${\rm U} (8)$ and a massless hypermultiplet
in the representations ${\bf 28} \oplus \overline{{\bf 28}}$ from the
untwisted sector, as well as 2 full massive hypermultiplets in the
adjoint representation for model 1 and in the ${\bf 28} \oplus \overline{{\bf
36}}$ for model 2 from the twisted sector. The doubling of the number of
hypermultiplets in the twisted sector is due to
the presence of the antisymmetric tensor \refs{\kakutor,\new} ,  
and to the fact that
the lattice sum ${\cal Z}_{2m+1}$ contributes twice\foot{This corrects
a mistake in \bg\ in the counting of
multiplicities for the open sector.} to each mass
level.
Since ${\cal A}$ and ${\cal M}$ are supersymmetric, the contribution
of the open unoriented sector to the 1-loop cosmological constant
vanishes identically.

As shown in \bg, one can derive this open unoriented spectrum by
modding out the supersymmetric $T^4 /Z_2$ orbifold compactification 
\refs{\ps,\bs,\gp} by the T-duality contained in $f$. This
operation is not always allowed. 
In fact, for orbifold compactifications in the presence of a quantized
antisymmetric tensor $B_{IJ}$, 
the transverse annulus amplitude involves the following terms \new
$$
\eqalign{
\widetilde{\cal A} \sim & \left[ 2^{r/2} \, (N+\overline{N}) \, \sqrt{v}
\,
+ \, {(D+\overline{D})\over \sqrt{v}} \right]^2 \, (V_4 O_4 - S_4 S_4
) + 
\cr
&\qquad + \left[ 2^{r/2} \, (N+\overline{N}) \, \sqrt{v}
\,
- \, {(D+\overline{D})\over \sqrt{v}} \right]^2 \, (O_4 V_4 - C_4 C_4
) \,,
\cr}
$$
coming from the torus amplitude \torb . 
In our case, it is crucial
that the compactification four-torus is an ${\rm SO} (8)$ lattice, since
at the ${\rm SO} (8)$ point
$$
\widetilde{\cal A} \sim \left[ (N+\overline{N} )+ (D+\overline{D})
\right]^2 (V_4 O_4 - S_4 S_4 ) + \left[ (N+\overline{N}) -
(D+\overline{D}) \right]^2 (O_4 V_4 - C_4 C_4 ) \,.
$$
Thus, we can mod out by T-duality, identifying the Neumann and Dirichlet 
charges\foot{In our conventions $N\equiv D$ ($N\equiv \overline{D}$)
corresponds to model 1(2).}, and obtain the amplitude
$$
\widetilde{\cal A} \sim (N+\overline{N})^2 (V_4 O_4 - S_4 S_4 ) \,,
$$
compatible with \tamm .

We now study the limiting behaviour for large and small radius. As
$R\to 0$, inspection of 
the transverse amplitudes reveals that new tadpoles arise due to
odd windings, that become massless. These new tadpoles receive a
contribution also from the transverse Klein bottle amplitude.
Trying to impose local tadpole cancellation, one then finds
$M-\overline{M} \not= 0$, in contrast with 
\tadpoles . 
This incompatibility can
not be resolved by adding Wilson lines as in the 
case of toroidal and supersymmetric
orbifold compactifications. Thus, for small $R$, the local
non-vanishing tadpoles induce 
linear divergences in the 5d gauge theory on the branes \ab.

The situation is different for $R\to\infty$. After doubling the radius,
the sums over momenta with alternating signs now vanish, while the others
give additional factors of two. In the closed unoriented sector this
leads to a massless spectrum consisting of 5 tensor multiplets and 16
hypermultiplets coupled to ${\cal N} =(1,0)$ supergravity in six
dimensions. The
interpretation of the open unoriented sector is more subtle. After
taking the limit $R\to\infty$
$$
\eqalign{
{\cal A} \sim & 2 \bigl[ 2 M \overline{M} (V_4 O_4 - S_4 S_4 ) + (M^2
+ \overline{M}^2 ) (O_4 V_4 - C_4 C_4 ) +
\cr
&\quad + 2 (M^2 + \overline{M}^2) (O_4 S_4 - C_4 O_4 ) + 2 M
\overline{M} (V_4 C_4 - S_4 V_4 ) \bigr]\,,
\cr}
$$
and
$$
{\cal M} \sim  - 2 (M +\overline{M}) (\widehat{O}_4 \widehat{V}_4 -
\widehat{C}_4 \widehat{C}_4 )\,.
$$
The factor of two reveals a doubling of degrees of freedom that
reflects a doubling of CP charges, $M \to (N,D)$. Then, 
consistency of the model leads to the following amplitudes
$$
\eqalign{
{\cal A} \sim & (2 N\overline{N} + 2 D\overline{D}) (V_4 O_4 - S_4 S_4
) + (N^2 + \overline{N}^2 + D^2 + \overline{D}^2 ) (O_4 V_4 - C_4 C_4
) +
\cr
&\qquad + 2 (2 N\overline{D} + 2 \overline{N} D) (O_4 S_4 - C_4 O_4 )
+ 2 (2 ND +2 \overline{N} \overline{D}) (V_4 C_4 - S_4 V_4 )\,,
\cr}
$$
and
$$
{\cal M} \sim - (N+\overline{N}+D+\overline{D}) (O_4 V_4 - C_4 C_4
)\,.
$$
All these results would be obtained directly starting from the
limiting torus amplitude.
The massless spectrum consists of ${\cal N}=(1,0)$ vector
multiplets in the adjoint of ${\rm U} (8) \otimes {\rm U} (8)$,with
hypermultiplets in the representations $({\bf 28};{\bf 1})\oplus
(\overline{{\bf 28}};{\bf 1})
\oplus ({\bf 1};{\bf 28}) \oplus ({\bf 1};\overline{{\bf
28}}) \oplus ({\bf 8} ; \overline{{\bf 8}})\oplus (\overline{{\bf 8}}
; {\bf 8})$. This is precisely the spectrum associated to the $T^4 /Z_2$
orbifold in the presence of a quantized $B_{IJ}$
\refs{\bs,\kakutor,\new}. The additional tensor multiplets take part
in a generalized Green-Schwarz mechanism for the cancellation of the
residual anomalies \ggsm .


\newsec{Supersymmetry breaking on the heterotic side}

An interesting question is whether the open sector of                        
the model discussed above remains supersymmetric at the non-perturbative
level. A way to address this question would be to find a heterotic dual.
Before breaking supersymmetry with the freely acting projection, the presence
of a quantized $B_{IJ}$ in the $T^4/Z_2$ suggests that a possible heterotic
dual should have reduced rank \refs{\newtor,\bgmn}. This is possible only for
the type I$^\prime$ description which is related to type IIA and has only one
tensor multiplet. Moreover, local tadpole cancellation requires to separate
the branes so that the resulting gauge group is 
${\rm U}(4)^2_{88}\otimes{\rm U}(4)^2_{44}$.
However, in the presence of supersymmetry breaking there is no perturbative
heterotic dual, since the action of the freely acting projection identifies
the D8 with the D4 branes on the type I$^\prime$ side.

Nevertheless, one can address the question of supersymmetry breaking in the
non-perturbative gauge sector of the type II model of \harv, by
analysing the partition function of its heterotic dual.\foot{Strictly
speaking, the type II model of \harv\ does not have points of enhanced
non-abelian symmetry because of the presence of the ${\rm NS}\otimes {\rm
NS}$ antisymmetric tensor in the ${\rm SO} (8)$ lattice. However, its
heterotic dual can be deformed to a point of enhanced gauge symmetry that we
discuss here.}
The starting point of the construction is the compactification on
the Narain lattice:
$$
\Gamma_{(5,21)}=\Gamma_{{\rm E}_8}\oplus \Gamma_{{\rm E}_8}\oplus
\Gamma_{{\rm SO}(8)}\oplus \Gamma_{(1,1)}(R)\, ,
$$
where the $\Gamma_{{\rm E}_8}$ factors refer to the affine $\widehat{\rm E}_8$ 
algebra, $\Gamma_{{\rm SO}(8)}$ denotes the four dimensional
compactification torus at the ${\rm SO}(8)$ symmetry
enhancement point and $\Gamma_{(1,1)}(R)$  denotes the
compactification on a circle of radius $R$. In terms of ${\rm SO}(8)$ 
and ${\rm E}_8$ characters, the partition function is then 
$$
{\cal Z}_{{\rm het}}^{(0)} = {1\over \eta ^4 \bar \eta^4} (V_8 - S_8 )
\, \overline{\chi}_{{\rm E}_8} \, \overline{\chi}_{{\rm E}_8} \,
\left[ |O_8 |^2 + |V_8|^2 +|S_8|^2 + |C_8|^2 \right] \, {\cal Z}_{m,n}
\,.
$$
The massless level contains the 5d ${\cal N}=4$ supergravity
multiplet coupled to vector multiplets with gauge group 
${\rm E}_8\otimes {\rm E}_8\otimes {\rm SO}(8)$.

The action of the non-supersymmetric orbifold gen\-erator $f_{\rm het}$ on the
heterotic side can be defined using heterotic--type IIA duality and
the adiabatic argument \harv. It is given by 
$$
f_{\rm
het}=\left[(1^5\,;\,-1^4\,,\,1)\,,\,(\delta^{5}\,;\,0^4,\delta)\,,\, 
S_2({\rm E}_8\otimes {\rm E}_8)\,,\, (-)^{F_{\rm L}}\right]\,,
$$
with $\delta$ denoting the action of a $Z_2$ shift
on the compactification lattice and $S_2$
the permutation of the two $\Gamma_{{\rm E}_8}$ lattices \harv. The
resulting partition function is 
$$
\eqalign{
{\cal Z}_{{\rm het}} = {1\over 2} {1\over \eta^4 ( q )
\bar\eta^4 (\bar q) 
} \Bigl[& (V_8 - S_8) (q) \,
(\overline{\chi}_{{\rm E}_8}\, \overline{\chi}_{{\rm E}_8})
(\bar q) \, \big[ |O_4 O_4 + V_4 V_4 |^2 + |V_4 O_4 + O_4 V_4 |^2 +
\cr
& \quad+
|C_4 C_4 + S_4 S_4 |^2 + |S_4 C_4 + C_4 S_4 |^2 \big]  \, {\cal
Z}_{m,n} (q ,\bar q) +
\cr
& + (V_8 + S_8) (q) \,\, \overline{\chi}_{{\rm E}_8} ( \bar q^2 )
\, \,
|O _4 O _4 - V_4 V_4 |^2
 \,\,
(-)^m {\cal Z}_{m,n} (q,\bar q) +
\cr
&  + {\textstyle{1 \over 4}}
(O_8 - C_8)(q) \,\, \overline{\chi}_{{\rm E}_8}
(\sqrt{\bar q}) \,\, 
| ( O _4 +  V _4) ( S _4 
+  C _4 ) +
\cr
&\quad+( S_4  + C _4  ) (
O _4 +  V_4 )|^2 \, \, {\cal Z}_{m,n+1/2} (q
,\bar q) +
\cr
& + {\textstyle{1 \over 4}}
(O_8 + C_8 ) (q) \,\, \overline{\chi}_{{\rm E}_8}
(-\sqrt{\bar q}) \,\, 
| ( O _4 - V _4 ) (S _4 + C _4 ) 
 +
\cr
&\quad+ (S_4 + C _4 ) (
O _4 - V_4 )|^2 \,\, (-)^m {\cal Z}_{m,n+1/2} (q
,\bar q) \Bigr]\, .
\cr}
$$

As a result of the shift on the circle, the twisted sector is massive,
while the $q$-expansion of the untwisted sector  
$$
(16+ 128 q+\ldots)(\bar q^{-1}+252 \bar q+\ldots)\,(-)^m {\cal
Z}_{m,n}\,,
$$
reveals a Fermi-Bose degeneracy for the massless states, because of the
absence of the $\bar q^0$ term.
It is interesting to study the contributions of the various massless
states to the vacuum energy. Expanding each term of the untwisted sector 
$$
\eqalign{
{\cal Z}_{{\rm het}}^{{\rm untw}} \sim & q^{-1/2} \bar q ^{-1} 
(1 + 4 \bar q) [(4+4) q^{1/2} - 2(2+2) q^{1/2} ] 
[1 +(248 + 248) \bar q ] \times
\cr
&\qquad \times [(1 + 6 \bar q )(1 + 6 \bar q) + 4\times 4 \bar q ]
 +
\cr
& + q^{-1/2} \bar q ^{-1} (1+4 \bar q) [(4+4)q^{1/2} + 
2 (2+2) q^{1/2} ] [1 + (248 - 248) \bar q ] \times
\cr
& \qquad \times [(1+6 \bar q) (1+6 \bar q) - 4\times 4 \bar q ] \, ,
\cr}
$$
one can see that the contribution of the massless states to the vacuum
energy vanishes for the following reasons. On the one hand, in the
``gravitational sector'', whose massless excitations are 
$\{ g_{\mu \nu}\,,\, B_{\mu\nu} \,,\, \phi \,,\, 4 A_\mu \}$,
$\{ A_\mu \,,\, 4 \phi \}$ and the spinors $\{ \psi_\alpha \,,\,
\bar\psi_{\dot\alpha}\}$
in the adjoint and $({\bf 4}\,,\, {\bf 4})$ representations of ${\rm SO} (4)
\otimes {\rm SO} (4)$, respectively, the $f_{\rm het}$ projection breaks
effectively supersymmetry, while preserving Fermi-Bose degeneracy.
On the other hand, the ``gauge sector'' is effectively supersymmetric. In
fact, the projection gives a plus sign to the space-time bosons $V_8$ and to
the diagonal combination of the two ${\rm E}_8$ factors and 
a minus sign to the space-time fermions $S_8$  and to the antisymmetric
combination of the gauge factors. As a result, a full
${\cal N}=4$ vector supermultiplet in the adjoint of ${\rm E}_8$
survives the orbifold projection. This property, however, holds only
for massless states. For instance, at the first mass level the
decomposition of the product of two adjoint representations, 
$$
{\bf 248} \otimes {\bf 248} = {\bf
1}_{{\rm s}} \oplus {\bf 248}_{{\rm a}} \oplus {\bf 3875}_{{\rm s}}
\oplus {\bf 27000}_{{\rm s}} \oplus {\bf 30380}_{{\rm a}} \,,
$$
reveals that bosons and fermions in the ``gauge sector'' appear 
in symmetric and antisymmetric representations, respectively, and thus do not
fit any more into supermultiplets. 

By heterotic--type II duality, one can then argue that supersymmetry remains
unbroken for the massless non-abelian gauge sector that arises 
non-perturbatively at singular points of K3 from D2 branes wrapped around
collapsing 2-cycles. This phenomenon is similar to the ``M-theory breaking"
of type I models with the direction of supersymmetry breaking transverse to
the D-brane \ads.


\newsec{Free-fermions and open strings}

An alternative approach to the construction of non-supersymmetric
vacua with vanishing cosmological constant is based on the fermionic
construction \abk . Using this approach, the authors of \st\ 
constructed a series of non-supersymmetric models in $D=4$ with 
Fermi-Bose degeneracy at each mass level. They found two different
classes of models related to asymmetric $Z_2 \otimes Z_2$ orbifolds. In
the first class the two $Z_2$ twists break ${\cal N} =(4,4)$
supersymmetry to ${\cal N}=(2,0)$ and ${\cal N}=(0,2)$,
respectively. In the second class supersymmetry is broken
to ${\cal N}=(2,0)$ for the first $Z_2$ and to ${\cal N} =(0,4)$ for
the second one. Although the
full $Z_2 \otimes Z_2$ model is not supersymmetric, each projection is
supersymmetric thus ensuring the vanishing of the cosmological constant. 
Only model I in \st\ is left-right symmetric and can therefore be modded out 
by $\Omega$ to construct open descendants.

The construction of the four dimensional $Z_2 \otimes Z_2$ orbifold
starts with the type IIB superstring
compactified on the ${\rm SO} (12)$ lattice whose partition function is
$$
{\cal T} = |V_8 - S_8 |^2 \Big[ |O_{12}|^2 + |V_{12}|^2 + |S_{12}|^2
+ |C_{12}|^2 \Big] \,.
$$
Following \abk , the $Z_2 \otimes Z_2$ projection can be defined by 
adding the following sets of periodic fermions 
$$
{\bf S}=\{ \chi^{I}\,,\, \omega^{I} \} \,,
\quad {\bf S}' = \{ \chi^I \,,\, y^I \} \,, 
\quad \overline{{\bf S}} = \{ \overline{\chi}^I \,,\, \overline{\omega}^I \}
\,, \quad \overline{{\bf S}} ' =\{ \overline{\chi} ^I \,,\, \overline{y} ^I
\}\,,
$$
to those that define the ${\rm SO} (12)$ lattice, where $I=1,\ldots ,4$
labels the coordinates of the internal $T^4$ on which the orbifold
generators act non-trivially.
In the orbifold language this translates into the generators:
$$
\eqalign{
f&=\Big[(-1^4, 1^2\, ; \,1^6)\,,\, (0^6\,;\, s^4,0^2)\,,\, 
(-)^{F_{\rm R}}\Big]
\cr
g &= \Big[ (1^6 \,;\, -1^4 ,1^2)\,,\, (s^4 ,0^2 \,;\, 0^6) \,,\,
(-)^{F_{{\rm L}}} \Big] \ ,
\cr}
$$
where $s$ is a $Z_2$ shift on the internal bosons.
As a result, the partition function of the orbifold model can be arranged 
into 64 characters. They can be generated from the identity
$$
\chi_{0,1} = V_4 O_4 O_4 O_4 O_4 + O_4 V_4 O_4 V_4 V_4  - S_4 S_4 V_4 O_4
V_4  - C_4 C_4 V_4 V_4 O_4 \,.
$$
by applying a modular $S$ transformation.
Besides $\chi_{0,1}$, the massless characters in the untwisted sector 
are: 
$$
\eqalign{
\chi_{0,2} &= V_4 O_4 V_4 O_4 V_4  + O_4 V_4 V_4 V_4 O_4 - 
S_4 S_4 O_4 O_4 O_4 - C_4 C_4 O_4 V_4 V_4  \,,
\cr
\chi_{0,9} &= V_4 O_4 O_4 V_4 V_4 + O_4 V_4 O_4 O_4 O_4 - 
S_4 S_4 V_4 V_4 O_4  - C_4 C_4 V_4 O_4 V_4 \,,
\cr
\chi_{0,10} &= V_4 O_4 V_4 V_4 O_4 + O_4 V_4 V_4 O_4 V_4 - 
S_4 S_4 O_4 V_4 V_4  - C_4 C_4 O_4 O_4 O_4 \,.
\cr}
$$
In the $f$-twisted sector, they are given by 
$$
\eqalign{
\chi_{f,1} &= O_4 C_4 O_4 O_4 S_4  + V_4 S_4 O_4 V_4 C_4 - 
C_4 V_4 V_4 O_4 C_4  - S_4 O_4 V_4 V_4 S_4 \,,
\cr
\chi_{f,2} &= O_4 C_4 V_4 V_4 S_4  + V_4 S_4 V_4 O_4 C_4 - 
C_4 V_4 O_4 V_4 C_4 - S_4 O_4 O_4 O_4 S_4 \,,
\cr
\chi_{f,3} &= O_4 C_4 V_4 V_4 C_4  + V_4 S_4 O_4 V_4 S_4 - 
C_4 V_4 V_4 O_4 S_4  - S_4 O_4 O_4 O_4 C_4  \,,
\cr
\chi_{f,4} &= O_4 C_4 O_4 O_4 C_4 + V_4 S_4 V_4 O_4 S_4 - 
C_4 V_4 O_4 V_4 S_4  - S_4 O_4 V_4 V_4 C_4 \,,
\cr}
$$
in the $g$-twisted sector by
$$
\eqalign{
\chi_{g,1} &= O_4 O_4 O_4 C_4 C_4  + V_4 V_4 O_4 S_4 S_4 - 
S_4 C_4 V_4 C_4 S_4  - C_4 S_4 V_4 S_4 C_4  \,,
\cr
\chi_{g,2} &= O_4 O_4 O_4 S_4 S_4  + V_4 V_4 O_4 C_4 C_4  - 
S_4 C_4 V_4 S_4 C_4 - C_4 S_4 V_4 C_4 S_4  \,,
\cr
\chi_{g,3} &= O_4 O_4 O_4 S_4 C_4  + V_4 V_4 O_4 C_4 S_4  - 
S_4 C_4 V_4 C_4 C_4  - C_4 S_4 V_4 S_4 S_4 \,,
\cr
\chi_{g,4} &= O_4 O_4 O_4 C_4 S_4 + V_4 V_4 O_4 S_4 C_4  - 
S_4 C_4 V_4 S_4 S_4  - C_4 S_4 V_4 C_4 C_4 \,,
\cr}
$$
and, finally, in the $fg$-twisted sector by
$$
\eqalign{
\chi_{fg,1} &= O_4 S_4 O_4 C_4 O_4  + V_4 C_4 O_4 S_4 V_4  - 
S_4 V_4 V_4 S_4 O_4 - C_4 O_4 V_4 C_4 V_4 \,,
\cr
\chi_{fg,2} &= O_4 S_4 V_4 C_4 V_4  + V_4 C_4 V_4 S_4 O_4 - 
S_4 V_4 O_4 S_4 V_4 - C_4 O_4 O_4 C_4 O_4 \,,
\cr
\chi_{fg,9} &= O_4 S_4 O_4 S_4 O_4  + V_4 C_4 O_4 C_4 V_4 - 
S_4 V_4 V_4 C_4 O_4  - C_4 O_4 V_4 S_4 V_4 \,,
\cr
\chi_{fg,10} &= O_4 S_4 V_4 S_4 V_4 + V_4 C_4 V_4 C_4 O_4 - 
S_4 V_4 O_4 C_4 V_4 - C_4 O_4 O_4 S_4 O_4 \,.
\cr}
$$
These massless characters appear in the partition function as follows
$$
\eqalign{
{\cal T}_{m^2 =0} =& |\chi_{0,1} |^2 + |\chi_{0,2}|^2 +
\chi_{0,9} \bar\chi_{0,10} +
\chi_{0,10}  \bar \chi_{0,9}
+ 
\cr
& + \left( \chi_{f,1} \bar \chi_{g,1} + \chi_{f,2} \bar \chi_{g,2} + \chi_{f,3}
\bar \chi_{g,3} + \chi_{f,4} \bar \chi_{g,4} + {\rm c.c.} \right) +
\cr
& + |\chi_{fg,1} |^2 + |\chi_{fg,2} |^2 
+ \chi_{fg,9} \bar\chi_{fg,10} + \chi_{fg,10} \bar\chi_{fg,9} \,.
\cr}
$$
The 4-dimensional massless spectrum resulting from the above expression 
yields 
$\{ g_{\mu\nu}$, $B_{\mu\nu}$, $8 A_\mu$,  $13 \phi$,  $16 \psi$,
$32 \phi$;  $16 \psi \} $ for the untwisted sector,
$\{ 32 \phi \,;\, 16 \psi \}$ for the $f$ and $g$ twisted sectors and 
$\{ 4 A_\mu \,,\,24 \phi \,;\, 16 \psi\}$ for the $fg$-twisted sector.

In order to construct the open descendants, one needs the characters
that appear in the partition function symmetrically in the holomorphic
and in the antiholomorphic sectors or combined with their conjugates.
Since in our case all characters are self-conjugate
\foot{${\rm SO}
(2n)$ characters are self-conjugate for even $n$.} the only
relevant contribution to the torus partition function reads:
\eqn\torus{
{\cal T}_{{\rm diag}} = \sum_{\alpha \in \{ 0,fg \}} \sum_{\ell=1}^{8}
|\chi_{\alpha , \ell} |^2 \,,
}
with
$$
\eqalign{
\chi_{0,3} &= V_4 O_4 V_4 O_4 O_4  + O_4 V_4 V_4 V_4 V_4  - 
S_4 S_4 O_4 O_4 V_4  - C_4 C_4 O_4 V_4 V_4 \,,
\cr
\chi_{0,4} &= V_4 O_4 O_4 O_4 V_4 + O_4 V_4 O_4 V_4 O_4 - 
S_4 S_4 V_4 O_4 O_4  - C_4 C_4 V_4 V_4 V_4 \,,
\cr
\chi_{0,5} &= V_4 O_4 S_4 C_4 C_4  + O_4 V_4 S_4 S_4 S_4  - 
S_4 S_4 C_4 C_4 S_4 - C_4 C_4 C_4 S_4 C_4 \,,
\cr
\chi_{0,6} &= V_4 O_4 C_4 C_4 S_4 + O_4 V_4 C_4 S_4 C_4 - 
S_4 S_4 S_4 C_4 C_4  - C_4 C_4 S_4 S_4 S_4  \,,
\cr
\chi_{0,7} &= V_4 O_4 C_4 C_4 C_4  + O_4 V_4 C_4 S_4 S_4  - 
S_4 S_4 S_4 C_4 S_4  - C_4 C_4 S_4 S_4 C_4 \,,
\cr
\chi_{0,8} &= V_4 O_4 S_4 C_4 S_4  + O_4 V_4 S_4 S_4 C_4  - 
S_4 S_4 C_4 C_4 C_4  - C_4 C_4 C_4 S_4 S_4 \,,
\cr}
$$
from the untwisted sector, and
$$
\eqalign{
\chi_{fg,3} &= O_4 S_4 O_4 C_4 V_4  + V_4 C_4 O_4 S_4 O_4 - 
S_4 V_4 V_4 S_4 V_4  - C_4 O_4 V_4 C_4 O_4  \,, 
\cr
\chi_{fg,4} &= O_4 S_4 V_4 C_4 O_4 + V_4 C_4 V_4 S_4 V_4 - 
S_4 V_4 O_4 S_4 O_4 - C_4 O_4 O_4 C_4 V_4 \,,
\cr
\chi_{fg,5} &= O_4 S_4 S_4 O_4 S_4 + V_4 C_4 S_4 V_4 C_4  - 
S_4 V_4 C_4 V_4 S_4  - C_4 O_4 C_4 O_4 C_4 \,, 
\cr
\chi_{fg,6} &= O_4 S_4 C_4 O_4 C_4  + V_4 C_4 C_4 V_4 S_4  - 
S_4 V_4 S_4 V_4 C_4  - C_4 O_4 S_4 O_4 S_4  \,,
\cr
\chi_{fg,7} &= O_4 S_4 C_4 O_4 S_4  + V_4 C_4 C_4 V_4 C_4 - 
S_4 V_4 S_4 V_4 S_4  - C_4 O_4 S_4 O_4 C_4 \,,
\cr
\chi_{fg,8} &= O_4 S_4 S_4 O_4 C_4  + V_4 C_4 S_4 V_4 S_4  - 
S_4 V_4 C_4 V_4 C_4 - C_4 O_4 C_4 O_4 S_4 \,,
\cr}
$$
from the $fg$-twisted sector. The $f$ and $g$-twisted sectors do not 
contribute to the Klein bottle, annulus and M\"obius strip amplitudes,
because they give rise to non diagonal contributions in the
partition function.

We can now proceed to construct the open descendants following
\refs{\cargese,\bs}. The direct channel Klein bottle amplitude is:
$$
{\cal K} = \sum_{\alpha \in\{ 0,fg\}} \sum_{\ell =1}^{8}
\chi_{\alpha,\ell} \,.
$$
The massless excitations of the closed unoriented sector comprise
the graviton, 4 abelian vectors, 54 scalars and 32 Dirac
spinors. 
The Klein bottle amplitude has ${\cal N}=2$ supersymmetry and therefore does
not generate any cosmological constant. 
In the transverse channel 
$$
\widetilde{\cal K} = 2^3\, \left( \chi_{0,1} + \chi_{0,2} +
\chi_{fg,3} + \chi_{fg,4} \right)\,,
$$
develops massless tadpoles, whose cancellation require the
introduction of open strings.

In principle, one is free to introduce in the Klein bottle amplitude
signs that are consistent with the crosscap constraint
\refs{\fps,\zero} (see also \zerop). 
The choice which includes eight positive and eight negative signs 
results in a model without an open sector, similarly to what was found in
\nopen . This follows from the unitarity of the $S$ matrix, that
implies that each character transforms into
$\chi_{0,1}$ with a positive sign, so that the transverse Klein
bottle does not contain the identity and has no IR divergences.
The only possible solution is an open descendant without 
open strings. 

Inspection of \torus\ allows for 16 different CP
charges. The transverse channel annulus thus reads
$$
\widetilde{\cal A} = 2^{-3}\, \sum_{\alpha\in\{ 0,fg\}} \sum_{\ell
=1}^{8} B^{2}_{\alpha ,\ell} \, \chi_{\alpha,\ell}\,,
$$
while the transverse M\"obius amplitude is
$$
\widetilde{\cal M} = 2 \Big[ \epsilon_{0,1} \ B_{0,1}\ 
\widehat{\chi}_{0,1} + \epsilon_{0,2} \ B_{0,2}\  \widehat{\chi}_{0,2} + 
\epsilon_{fg,3} \ B_{fg,3}\  \widehat{\chi}_{fg,3} + \epsilon_{fg,4}\ 
B_{fg,4}\ \widehat{\chi}_{fg,4} \Big]\,,
$$
where the boundary-to-boundary coefficients $B_{\alpha ,\ell}$ are 16
orthogonal combinations of the CP multiplicities and $\epsilon_{\alpha
,\ell}$ are signs. The solution of the inhomogeneous tadpole conditions
$$
\eqalign{
B_{0,1} &= \sum_{\ell =1}^{8} n_{0,\ell} + \sum_{\ell=1}^{8}
n_{fg,\ell} = - 8 \, \epsilon_{0,1} \,,
\cr
B_{0,2} &= \sum_{\ell =1}^{8} n_{0,\ell} - \sum_{\ell=1}^{8}
n_{fg,\ell} = - 8 \, \epsilon_{0,2} \,,
\cr}
$$
then requires that the $n_{fg}$ charges vanish identically and fixes
the signs $\epsilon_{0,1} = -1 = \epsilon_{0,2}$ in the M\"obius 
strip amplitude.

In the direct channel, the annulus and M\"obius strip amplitudes
can be cast into  ${\cal N} = 2$ supersymmetric extended characters, thus
ensuring the vanishing of the cosmological constant also in the open
unoriented sector. Introducing a minimal set of charges, these amplitudes
are 
$$
{\cal A} = 2 n \overline{n} (\chi_{0,1} + \chi_{0,2} + \chi_{fg,3} +
\chi_{fg,4} ) + (n^2 + \overline{n}^2 ) (\chi_{0,3} + \chi_{0,4} +
\chi_{fg,1} + \chi_{fg,2} ) \,,
$$
and
$$
{\cal M} = (n+\overline{n} ) (\widehat{\chi}_{0,3} +
\widehat{\chi}_{0,4}) +  (n-\overline{n}) (\widehat{\chi}_{fg,1} +
\widehat{\chi}_{fg,2}) \,.
$$
The corresponding massless excitations are an 
${\cal N}=2$ vector multiplet in the adjoint representation of
${\rm U} (4)$ and one hypermultiplet in the ${\bf 10} \oplus
\overline{\bf 6}$ representations. The reduction of the rank of the CP
gauge group is due to the presence of a rank four antisymmetric tensor
in the definition of the ${\rm SO} (12)$ lattice \toroidal .


\newsec{Conclusion}

In this paper we studied open descendants of non-supersymmetric type
IIB compactifications with zero cosmological constant. The
construction of the parent closed string theory is based on a freely
acting orbifold that resembles an asymmetric Scherk-Schwarz
deformation. Whereas supersymmetry is broken in the bulk, the open
sector remains supersymmetric at all mass levels.
An interesting open question concerns the radiative corrections
induced by the
non-supersymmetric bulk in the supersymmetric open string spectrum. In
particular, it would be interesting to understand
the magnitude of the induced mass splittings, and whether
these splittings preserve the property of the vanishing vacuum energy.

A possible application of the constructions discussed here is to models with
a low string scale and supersymmetry breaking by large dimensions
\refs{\a,\low,\kt}.  An immediate
limitation, however, is that these constructions 
allow only one free internal
dimension, since four of them are fixed at the fermionic point, while 
the fifth
one determines the scale of supersymmetry breaking in the bulk. One
possibility within this limitation is, for instance, to take the
compactification scale of the fifth dimension close to the string scale at
intermediate energies, in such a way that the (gravitationally) induced mass
splittings on the brane are of the order of a TeV. This may be possible by
adjusting the size of the remaining (free) transverse dimension, that can be as
large as a millimeter, implying a string scale as low as $10^8$ GeV.
Furthermore, in order to avoid large (linearly divergent) corrections to the
effective field theory, one should impose local tadpole cancellation, that
leads to additional constraints in model building \ab. New possibilities,
however, may arise if some of the four 
fixed internal radii are liberated in more
general constructions. 

In the context of type II theories with low string
scale \ap, it is also interesting to study the closed string models with
vanishing vacuum energy  at special points of moduli space, where
non-perturbative gauge symmetries appear. In particular, one should understand
the effects of supersymmetry breaking in the non-perturbative gauge sector of
the theory. In this work, we argued that supersymmetry remains unbroken for
the massless excitations to lowest order, by analysing the partition function
of the heterotic dual. These models are particularly
attractive since the induced non-perturbative cosmological constant will be
exponentially suppressed in the weak coupling limit.

\vskip 36pt

{\bf Acknowledgment.} We are grateful to M. Bianchi, J.A. Harvey, 
L.E. Ib\'a\~nez, E. Kiritsis, H.P. Nilles, B. Pioline and
Ya.S. Stanev and in particular to A. Sagnotti 
for discussions and comments.
This work was supported in part by EEC under the TMR contract 
ERBFMRX-CT96-0090.

\listrefs

\end